# Ten Incredibly Dangerous Software Ideas

## George A. Maney

**Abstract**

This is a rough draft synopsis of a book presently in preparation. This book provides a systematic critique of the software industry. This critique is accomplished using classical methods in practical design science.

## Introduction

Today the software industry is isolated, inward looking, insular, inbred, and inertial. Ideas from other design disciplines are irrelevant. They are irrelevant because practical software science is perfect.

Practical software science, for all intents and purposes, is indeed perfect. It is demonstrably perfect for exactly the sort of handcrafted pattern programs produced everywhere today. In this sense it is optimized, overspecialized, orthodox, and ossified yet also obsolescent.

For just this reason complacency is now a software industry institutional crisis. New ideas are permitted only when they don't seriously question the prevailing perfection of practical software science.

Even so, much remains to be done with practical software. So far the software industry has just begun to scratch the surface of customer value opportunities. The mother lode of these opportunities remains out of reach today.

In this strategic sense software science is stuck, stalled, and stymied. Some limited tactical software science progress continues to be made.

This progress mostly involves design possibility rather than design productivity. This progress occurs variably in myriad immature and incipient software industry product segments.

The last major strategic advances in software science were made in the 1980s. These include structured software design, simultaneous software design, and the super-specialty of software engineering.

The last minor strategic advances in software science were made in the 1990s. These involve Internet informatics specializations of middleware, messageware, and human-computer interface modeling.

So software science clearly needs a kick-start. Hence the time has come for some incredibly dangerous software ideas. These ideas are necessary in order to maintain the long-term progress and profit potential of the software industry.

Ten incredibly dangerous software ideas are presented here. In some way each is a significant threat to the established order of the software industry. These are the highest heresies available anywhere today in the software industry.

None of these ideas make any sense from the software industry perspective. Even so, all of these ideas make perfect sense elsewhere else in the world of practical design.

These dangerous ideas have always made perfect sense from the perspective of other major design disciplines. Ideas that are heretical in software design have long been habitual in all other major design disciplines.

The reason for this involves practical design science. Practical design science is always an important consideration whenever practical designs are done.

Software stands out as the most design-intensive product of all. So every strategic and tactical issue in the software industry involves design science in some significant way.

For this reason practical design science provides a unique unifying software industry perspective. Many seemingly unrelated issues are seen to be strongly related. Much that has always been seen as exotic, eccentric, esoteric, or enigmatic is easily explained.

Design science is and always has been different in the software industry. Software design science is a lot different rather than just a little different. This polarity is all about patterns.

Pattern design science and principled design science are the two primary varieties of practical design science. These are mostly mutually antithetical.

Moreover, patterns and principles don't mix well in practical design. It is traditional in design science to say that "principles poison patterns and patterns poison principles."

Today the software industry is the last remaining refuge of the professional pattern designer. Pattern design has been extinct for decades outside of the software industry. In these other design disciplines principled design is obligatory and pattern design is outlawed.

The primitive sort of pattern design done handcrafted pattern programming today is profoundly antique. Suppose that we could somehow magically materialize the plant and product of a contemporary pattern programming project. The result would resemble nothing so much as the controlled chaos of an automotive craftwork shop *circa* 1900.

What are the implications of patterns vs. principles in the software industry? This design science question is the genesis of the ten incredibly dangerous ideas considered hereafter.

## Practical Information is Not Computation in Disguise

This is absolutely the most dangerous software idea of all, bar none. Practical software production will cease if this idea takes hold. The software industry will simply grind to a halt.

The idea that practical information is computation in disguise is the starting point for all practical software production today. This idea is the cornerstone of pattern programming design science.

Pattern programming as practice today is best understood as a highly specialized kind of design science. This is a pattern design science. It is a unique formalist kind of pattern design science.

A fundamental tenet of classical design science is the distinction between pattern and principled design. This are the two primary sorts of practical design science know in human history.

Today pattern design science is the exclusive province of the software industry. All other practical design disciplines abandoned patterns for principles many decades ago. The simple reason for this is that principles are far more powerful and precise than patterns.

Today pattern design is effectively outlawed everywhere but in the software industry. High standards of principled design are mandated and maintained in all other design disciplines. Principled design is obligatory in general management, the principled professions, and the statefull sciences.

Why is pattern design science still pervasive in the software industry? The reason is that principled practical software production is far beyond mortal means.

All practical design is topical. Today the world of practical design encompasses myriad thousands of topics and myriad millions of targets.

Design science tells us that design pattern provide poor reusability. So each new software project needs a specialized *de novo* and disposable pattern design space.

The genius of pattern programming is patterns. Pattern programming is possible only because of a vast diversity of patterns.

So topical specialization of pattern software is utterly impractical. Other design disciplines seek total topicality in design science. Software design strives for absolute atopicality. This is why software design models, methods, and mindsets are always atopical.

Atopicality is the ultimate design science compromise. Atopical software design is workable for all kinds of practical software applications. Even so, it's not especially well suited for any particular kind of software application.

Atopical design requires designers to look at practical information as computation in disguise. This is the absolutely atopical way of looking at practical information. This is the starting point for atopical software modeling approaches such as entity-relationship and object-oriented approaches.

Approaches of this sort dominate practical software production today. These are demonstrably engineering and economic optimal for handcrafted pattern program production.

## Practical Information is Civilization in Disguise

This is an incredibly dangerous idea because it's the other way of looking at practical information. Software customers understand practical information as civilization in disguise. This view of practical information is a total mystery in the software industry today.

Software customers find the notion of practical information as computation in disguise to be utterly incomprehensible. They don't understand computation. They couldn't care less about computation. They only understand the practicalities of civilization and practical information.

In the software industry practical software is always seen as an end in and of itself. Yet software customers never see software in this way.

Software customers always see software as a means to and end. That end is always automating and animating practical information systems.

These are the information systems that underlie the vast diversity of mass modeling and mass management systems in service today. Nearly all practical information flows in modern industrial civilization involve these mass information systems.

Just exactly what is practical information? What are the infodynamics of practical information flows like?

Today in the software world nobody knows. The software industry has successfully ignored practical information and practical infodynamics until now.

All aspects of practical information are irrelevant in pattern programming. Here there is one and only way to look at information. This is as computation in disguise. All other ways are distracting and thus dysfunctional.

How do practical people use practical information in practical pursuits? This is clearly an interesting and important question. It is ultimately the key to understanding practical software customer value. Even so, it is not a question that the software industry knows how to ask today.

You may well be skeptical. How can software be used to automate practical information systems without a good understanding of practical information? How does this make sense?

In pattern programming this makes perfect sense. The reason is that pattern programming always involves a systematic trivialization of practical information.

Pattern programs always start with principled practice models provided by software customers. In pattern programming these models are always restated as pattern practice models.

This is unavoidable. A pattern program is a structure of software patterns. Programmers must select and structure these patterns.

This is an analogical matching process. Software patterns are chosen and combined to maximize the analogical match to subject matter side patterns and pattern structures. Patterns beget patterns in pattern programming.

Subject matter side source restatement is a terrible and traumatic thing to put any principled practice model through. The resulting pattern source model is never more that the merest shadow of the starting source model.

All symbolic specifications, including all pattern programs, are ultimately structures of details. In pattern programming subject matter detail structures are decimated. Details are divorced from the institutional information contexts that provide most all information quality. Thus most of the source model meaning is lost in translation from subject matter to software.

Why do we abuse innocent information in all these awful ways? The answer is that we must. Otherwise pattern programming would be entirely impractical.

Every software project manager knows that subject matter trivialization is unavoidable. Skillful trivialization is essential to survival and success in pattern program production.

Trivialization is a tricky business. This often amounts to deciding which parts of the information baby to throw out with the bathwater. Substantial skill building is required to master the arts of successful trivialization.

Yet without a suitable subject matter trivialization software will never see the light of day. Pattern programming is profoundly inefficient, ineffective, and inexpedient. Human intellectual gifts and graces are limited. So without a clever trivialization converting subject matter to software is simply beyond mortal means.

How is source model trivialization accomplished? Just about everything done in practical programming contributes to trivialization in one way or another.

All software modeling methodologies involve systematic trivialization. Methodologies such as entity-relationship and object-oriented modeling methodologies are dominant today. This dominance is largely because they provide robust and reliable means of achieving successful systematic source model trivializations.

## Computation is Not a Universal Science

The starting point for all of theoretical computer science is universal computation in the Turing sense. Universality in this sense has a very highly specialized mathematical meaning. This mathematical meaning really isn't very interesting for most practical purposes.

In pattern programming design science computational universality has a deeper meaning. Here the notion is that computation is universal in a broader sense. This other notion of computational universality has always been an important theme in computer science.

A science is any systematic way of understanding order. A universal science is a unified way of understanding the entirely of observable and occult materialist, mathematical, and metaphysical order. The search for universal science starts with the origins of philosophy both in eastern and western civilization.

The conjecture that computer science is a universal science is a recurring theme in computer science and software science research. Sometimes this is merely a minor theme, but all too often it is the major theme. Some computation as universal science conjectures include:

- Cosmology is computation in disguise.
- Cognition is computation in disguise.
- Civilization is computation in disguise.

Well, so what? What's wrong with these conjectures? The answer is twofold:

- It doesn't advance our understanding of computation.
- Is doesn't advance our understanding of cosmology, cognition, or civilization.

The goal of good research is to systematically clarify scientific understanding. Viewing computation as universal science in research is always contrary to this goal. It always systematically confuses and clouds scientific understanding.

This sort of systematic confusion is that starting point for understanding practical information as computation in disguise. This necessarily presumes that computation is an underlying universal design science. Only as a universal science can computation claim to subsume the myriad utilitarian design sciences of diverse principled practice.

Today, after fifty years of exceptional effort and expense, we understand a lot about computation. We understand computation far better than we understand most other things. There is a vast arsenal of alternative formalist models, methods, and mindsets for understanding computation.

Even so, we don't understand practical information today very much better than we did fifty years ago. So in this sense computation has failed as a universal science. Understanding computation has not succeeded as a substitute for understanding practical information.

Some will claim that success in this sort of substitution lies in the future. In this view understanding practical information in terms of computation is inevitable. Many years, or perhaps even decades, of further research is required.

Yet this is unlikely. Even after fifty years of exploring computation there is no empirical evidence that computation has any non-trivial materialist meaning. The only verifiable meaning of computation so far is mathematical. The practical design sciences underling practical information are always materialist rather than mathematical.

So it would seem that there is only one remaining way to understand practical information. This way is to investigate practical information and the underlying practical design science. Even so, investigating practical information is problematic in the software industry today.

The reason is that in the software industry today information and computation are synonymous. These don't seem to make sense as separable issues. This confusion is a natural consequence of looking at information as computation in disguise.

So studying practical information *per se* can only be seen as an inappropriate and incongruous idea. Studying computation and studying practical information must be exactly the very same thing.

The available understanding of practical information is a general bottleneck in delivering software industry customer value. This has been a bottleneck for a few decades. It will remain a bottleneck for decades to come.

It has been at least a decade since the available understanding of computation was any sort of general customer value bottleneck. Even so, the software industry still operates as if computation still is and always will be the big customer value bottleneck.

This will continue as long two dogmas remain *de rigueur* in the software industry. The first dogma is that practical information is computation in disguise. The second dogma is that computation is a universal science.

## Programming Cannot Be Advantageously Automated

The software industry has always envisioned pattern programming design automation as the eventual way forward in pattern program production. Yet pattern programming automation is far from a practical reality today.

The simple reason for this is that pattern programming cannot be advantageously automated. It cannot be automated by any means we understand and use today. Nor can it be automated by any alternative means.

The reason for this is simple. Today there are no workable solutions to precision practice modeling in pattern program production. Precision pattern practice modeling is intrinsically impossible. So there is simply nothing for automation to work with.

Other design disciplines carefully cultivate precision principled practice modeling. These invest heavily in principled design systems including design modelware, design automation, and design factories.

Thus other design disciplines are progressing by leaps and bounds in design productivity and possibility. Meanwhile the software industry struggles mightily for the most minuscule of design progress advances.

All manual design involves lots of wasted mental motion. Pattern programming uniquely involves vast amounts of wasted mental motion.

The wasted mental motion disadvantage of pattern design science is extreme. This disadvantage with respect to equivalent principled design science is usually several orders of magnitude.

Design automation is always the answer to wasted mental motion in manual design. All practical design automation advantage ultimately flows from reducing wasted mental motion in design. In a perfect design automated world no manual design decision would ever be made more than once in all of history.

All forms of practical design automation in service today employ principled design science modelware. Advantageous pattern design automation has never been demonstrated in any design discipline.

There is no such thing as universal design automation. Each topical sort of design automation relies on a combination of specialized topical techniques.

Each specialized sort of topical design automation technique provides particular advantage. This particular advantage results from by some specific reduction of wasted mental motion.

There are many broadly and narrowly specialized design automation techniques. Automation engineers devise, design, and develop these specialized techniques.

All of these techniques are always information-intensive. They need vivid and varied application information to work with. So they only work well with information maximal application modeling. Automation advantage increases in proportion to application model information-intensity.

This stands to reason. Design automation is not magic. Design automation needs design information to automate. Less design information means less design automation. Less design automation means less design automation advantage.

Pattern software application modeling is always aggressively information minimal. In fact whole huge pieces of the application model puzzle are missing in routine pattern programming applications.

It is for exactly this reason that pattern software design cannot be advantageously automated. There just isn't any interesting information for design automation techniques to work with. Thus there is no available design automation advantage.

This unique lack of interesting information is intrinsic to pattern design science. This explains exactly why advantageous pattern design automation has never been demonstrated in any practical design discipline.

# Pattern Programming is Not Perfectible

It has always been assumed in the software industry that pattern programming is perfectible. In this view the power of pattern design science can be pushed to parity with principled design science. Thereafter this power can be pushed onward to perfection.

Understanding this perfectibility assumption starts with the notions of universal language and universal logic. Every digital computer comes with universal language and universal logic as standard equipment. It's hidden somewhere under the covers of every computer we use everyday.

This is a consequence of universal computation. In this very particular mathematical sense every digital computer, within finite limits, is a universal computer.

Thus pattern programming must be perfectible. All this requires is unlocking and unleashing the power of universal language and universal logic. This just entails finding the right algorithms and algebras for any given application.

The problem is that the selection of algebras and algorithms is unbounded. Moreover, universal computation doesn't provide any help in selecting suitable algebras and algorithms. It merely tells us that these can be automated and animated once a suitable selection has been made.

In just this same way a paper and pencil are universal. You can write or draw anything that can be written or drawn with a pencil and paper. Even so, the paper and pencil are never very helpful in deciding what to write or draw.

Universal computation tells us that digital computers can be used to automate and animate all kinds of practical information. Even so, nothing in computer science or software science provides a canonical means of constructing the software necessary for this purpose.

Today successful software production depends on pattern programming craftwork cleverness. Programmers master vast stocks of software precedent and prototype patterns. Software design draws upon these stocks to structure a suitable software solution. This requires many hundreds of chess games worth of approximate and analogical reasoning.

This is the state of the art in pattern programming today. It works for exactly how software is produced today. Even so, the software industry can't even begin to produce most of the software that its customers really want in this way.

There are two issues in the perfection of pattern programming. These are the perfection of software style and the perfection of software substance.

Perfection of style is an effort that continues everywhere everyday in the software world. Endless pattern fads, fashions, factions, and feuds have always been endemic in the software world. These are a direct result of wasted mental motion. These are all driven by the hope of more powerful patterns.

The hope is that these patterns might eliminate much of the wasted mental motion involved in pattern programming. The hope is that these might radically improve pattern programming productivity.

These hopes are all in vain. It is exactly extreme pattern diversity that makes pattern programming possible, practical, and profitable. More pattern diversity is always preferable to less pattern diversity.

So there is no hope of finding perfect platonic software pattern frameworks. Moreover, we know with that we can never hope to discover more powerful software languages than those available today. We know this with absolute mathematical certainly.

Perfection of substance has always been the province of formalist software science research. Here the notion is that unlocking and unleashing the power of universal language and universal logic is the key to pattern programming perfection.

Here the hope is that some brilliant unanticipated formalist software science advance. This advance will unlock the power of universal language and universal logic. This advance will achieve the perfection of pattern programming.

This sort of radical formalist breakthrough is a software science holy grail. Breakthroughs of this sort have long been sought by many in the software science research world.

There has never been any shortage of promising candidate formalisms. Yet none of these formalisms, so far at least, seem likely to provide the surprise being sought.

Formalist software science has advanced rapidly over the course of the last quarter century. Much of this advance has been driven by the search for breakthrough formalisms. Yet this very advance has largely been its own undoing.

Due to this advance we now understand much about the intrinsic limitations of formalist software science. These limitations were unsuspected a quarter century ago. They seem not to bode well for spectacular software science surprises.

For this reason the situation in software science research is very different from that prevailing a quarter century ago. At that time the prospects for various kinds of spectacular software science surprises seemed bright indeed.

Today there is still a vast volume and variety of important work remaining to be done in formalist software science. Yet the prospects that this work will produce a spectacular software science surprise now appear very dim indeed. It is no longer reasonable to expect that this work will lead to the perfection of pattern programming.

In summary neither the style nor the substance of pattern programming is in any danger of perfection at present. Neither design science first principles nor practical programming experience support the notion that pattern programming is perfectible.

## There is Another Way to Produce Practical Software

There has always been and always will be exactly one way to produce practical software. That way is pattern programming.

What if there is actually one other way? What if this other way has the potential for extreme advantage? What if this other way doesn't involve programming or programmers? What if this other way involves a totally new and very different kind of software design science?

There has never been any shortage of radical new software production approaches. These have come and gone continuously since the earliest days of software.

Most commonly these involve some formalist variety of software science magic. Otherwise they are mostly exotic new styles of pattern program models, methods, or mindsets.

What about using exactly the same design science for practical software production that our customers use? This idea has never been seriously considered in the software industry.

The reason is that this idea entails the view of information as civilization in disguise. So the very idea is simply inconceivable in the software industry.

Design science tells us that it must be possible to produce software in this way. Even so, that's all that design science has to say about the matter. All the rest is an automation science challenge.

Why produce practical software this way? Just because most all software customers already have mature principled design sciences.

The principled systems, standards, spaces, and stocks are ready to be automated and animated. These are the ideal feedstock for producing principled practical software.

These powerful and precise principled models taken together are the sum and substance of potential software industry customer value. Using these directly to produce practical software is the ideal way to exploit all available software customer value.

One problem is that software customers have lots and lots principled design sciences. There are myriad millions of topic and target specialized varieties of principled practical design sciences.

These varieties support the entirely of the practical modeling legacy. The replacement value of this legacy is probably somewhere around one hundred trillion dollars. Something like five trillion dollars is spent on practical principled modeling every year worldwide.

The good news is that this legacy is ready and raring to go. It is useable as is for principled software production.

There is no need to redundantly restate this practical legacy in terms of patterns. Any pattern approach to practical programming automation would require exactly this sort of redundant restatement.

Practical principled software production mass custom production can be accomplished with proven automation technologies. It can be achieved by adopting, adapting, and applying various existing commercially mature and manageable automation technologies. These are trusted, time-tested automation technologies with track records.

Note that nearly all digital computer hardware has been designed by automated means using principled design science methods for about twenty years. The cumulative economic and engineering benefits of this approach so far amounts to five or six orders of magnitude.

In practical terms these benefits mostly involve radical increases in design productivity and design possibility. Without these benefits the personal computer, the Internet, and the cell phone as we know these today would not exist. Moreover, the digital semiconductor industry would be merely a minute faction of its current substantial size.

The digital design automation community has already solved all the hard problems in advantageously mass producing custom complex computational models. If we can produce hardware this way then we can certainly produce software this way.

We known that this is true because Turing equivalence tells us that hardware is just frozen software. Moreover, we can expect very much the same sort of extreme economic and engineering advantages.

This is not to say that existing digital design automation products should be pressed into in software production service. In fact this has actually been done on odd occasions. Even so, this is not a sufficiently general starting point for advantageous general purpose practical software production.

The key point here is that all the primary pieces of the principled software mass production puzzle are available somewhere in the automation software world today. So putting this puzzle together involves substantial design challenges but no real basic or applied research challenges. The design challenges involved are entirely, if not always easily, manageable.

In particular no formalist software science breakthroughs are needed. No spectacular software science surprises are needed. We don't need square the circle.

## Artificial Intelligence Isn't The Way to Smarter Software

What exactly is artificial intelligence (AI)? Even AI wizards have trouble answering this question. AI is many things to many people. For present purposes AI is about "smarter" software in general. Today most software isn't very smart. Nearly all practical computing applications involve high information quantity informatics. High quality informatics remains mostly infeasible.

This has always been a particular disappointment to practical software customers. These customers have always hoped for much "smarter" software that the software available today. Yet there is only a vast value vacuum where this smarter sort of software ought to be today.

So-called "hard" AI is the science fiction interpretation of AI. This is about making computers in our own image.

Hard AI is about using computers to simulate something resembling general human intelligence. It is about simulating various aspects of human perceptions and psychology. It is about simulating human cognition and cogitation.

Hard AI can be considered a brilliantly successful failure today. Generic human intelligence simulation is still a dream of a distant future. Perhaps it always will be such a dream.

Yet both hard AI and the more recent "soft" specializations of AI have been a seminal source of all sorts of interesting and useful pattern models. Many of these have found widespread commercial applications so far and more are on the way.

What does "smart" mean in practical software? Practical people use computers as tools. In this roles software serves as a mechanism for modeling and managing. Software enables us to use computers as tools for making, moving, and manipulating practical information models.

Today computers are still very rudimentary tools for this purpose. This is true because available information models are trivial. They are trivial only because pattern programming must always systematically trivialize these models.

Can't this trivialization be eliminated? Yes and no. Trivialization can be eliminated with novel principled approaches to practical software production.

Trivialization cannot be eliminated with any reasonable pattern programming approach. This is likely the most important lesson that can be learned from AI.

From a software standpoint AI can be seen heroic pattern modeling. Here the notion is that it might not be necessary to trivialize practical information source models in pattern program production.

AI has demonstrated that this trivialization can be reduced, but not radically reduced. The heroic pattern modeling required is heroic indeed. This extreme modeling exertion stretches the feasible limits of pattern programming past the point of pathology.

The result is smarter pattern software, but software that is not scaleable or safe. This lack of scalability and safety in heroic pattern programming is intrinsic. Thus hard AI can never be the ultimate answer to smarter software.

Today computer tools are not nearly as useful as they could be. They aren't very good at reducing wasted mental motion. They aren't very good at coping with complex cases or curious contingencies. They can't solve routine repetitive problems well. They can't dig up practice precedents or prototypes suitable for a given problem at hand.

Can't AI do this? Perhaps rarely in the research lab but not routinely in real life. Yet there is another kind of serviceable software that actually does these sorts of things all the time. This is high end design automation software.

There are a few dozen major and a few hundred kinds of captive and commercial high-end design automation today. These are specialized to narrow areas of enterprise, economic, and engineering practice. The cost of ownership of this sort of software runs to thousands or tens of thousands of dollars per year per seat.

Today design automation software is easily the smartest sort of software around. This is not to say that there isn't some remarkably clever AI software in service today. IBM™'s Deep Blue™ chess computer famously defeated world chess champion Garry Kasparov in 1996. The feat resulted from classic methods of heroic heuristic hard AI modeling.

Yet Deep Blue™ is not nearly the smartest software at IBM. IBM has many state-of-the-art design factories for mechanical, electronic, packaging, and other sorts of practical design.

Much of the software in these factories puts Deep Blue™ to shame. This smarter software is design automation software. It is not in any sense artificial intelligence software.

Smart software is ultimately about high information quality and high information safety. These cannot be reliably achieved purely be means of pattern programming design science.

Isn't high-end design automation software pattern programmed? In fact most of this sort of software is a hybrid of patterns and principles.

Smartness is achieved by incorporating elements of principled design science modelware. The software that works against this modelware is heroic pattern programming.

This demonstrates that elements of principled modeling provide a substantial smartness advantage. Totally principled software and modelware design science automation is the ultimate answer to smarter software.

What do practical software customers want? They want to advantageously automate their existing practical modeling systems, modeling standards, modeling spaces, and modeling stocks. Moreover, they want to preserve the entirety of existing information quality and information safety. Ultimately this is the sum and substance of all practical software customer value. This is all that practical software customers have ever wanted and all that they ever will want.

So far the software industry has managed to automate only a tiny fraction of the installed base of practical information systems. These mostly provide high quantity, low quality automation. They mostly simulate copyists, clerks, calculators, and couriers.

This sort of software can work with model media but not much with meaning. In high-end design automation today practical model meaning is routinely captured, composed, checked, compared, contrasted, collected, completed, critiqued, and coached.

This is not super-intelligent software. It is state-of-the-art automation software with principled design science modelware.

Why is this modelware so smart? Because it is not pattern programming modelware. So it is not systematically trivialized. It preserves much more of the original subject matter source detail structure.

In a sense smarter software is simple. Merely avoid systematic trivialization of subject matter source models. Software customer model stocks already have all of the information quality and information safety available. These stocks cannot be improved upon. These stocks are the sum and substance of practical institutional intelligence.

So the trick is to avoid throwing this information value away. Unfortunately this is impractical with pattern programming. In pure pattern programming trivialization is unavoidable. Thus so alternative sort of software production is required for smarter software. This alternative must employ principled design science software production.

Is this really the answer to smart software? Don't software customers want super-intelligent software? Don't they want automated expertise and eloquence? Don't they want empathetic software with extreme emotional intelligence? Don't they want companionable and clubbable computers? Don't they want sassy and savvy computes?

Apparent they don't. Practical software customers have never expressed any disappointment over the shortage of science fiction software.

The next step in practical computing is far smarter practical modeling tools. Model automation is always about maximizing the productivity and possibility of natural intelligence. This is automation of institutional intelligence. This is the antithesis of artificial intelligence.

Automation of institutional intelligence is what practical people everywhere want and need. This is sort of smarter software that will support them in practical pursuits.

Artificial intelligence is what practical people everywhere fear and dread. The certainly do not want smarter software that will supplant them in practical pursuits.

## Practical Language is a Technology *Per Se*

The ultimate in modeling technology is pencil, paper, and practical prose. Primary varieties of practical prose include verbal prose and visual prose.

Practical prose is demonstrably necessary and sufficient to capture the entire mechanistic practical meaning of any practical artifact. This includes all mundane as well as all managerial artifacts.

All practical software programs, in one way or another, are surrogates for pencil, paper, and practical prose. The source models for are practical programs are provided in practical prose.

How do we manage to turn prose into programs? Today there is no practical understanding of this process.

Software technical literature can only provide vague or vacuous accounts of this process. These accounts are assembled from psychological and philosophical abstractions.

Yet somehow explicit practical principled prose meaning is converted to implicit practical program meaning. Looking at practical language as a technology *per se* is the starting point for understanding this conversion.

Practical language as technology is not the only way of looking a practical language. This view opens up all kinds of new ways to model and mechanize practical language.

Even so, understanding human language is incredibly complex. Ultimately multiple views of language will be required to crack this complexity. Practical language as technology can never be more that just one of many ways of looking at language.

The notion of practical language as a high and hard technology in its own right is not new. This was a popular and practical notion in parts of the old industrial science movement.

The notion of strong underlying structure in the substance of practical language is hardly novel. The sophists of Greek antiquity made a major point of this.

Today we understand technology well enough to understand that this rich, regular, and redundant structure is a technostructure.

Any artifact, including practical language, with an observable technostructure can be hacked. Hacking practical language is the ultimate high technology trophy hack.

This is true because practical language is the ultimate technology. It is the only technology that is always necessarily the starting point for every other technology.

Moreover, practical language is the one absolutely indispensable technology in any human civilization. It is the only technology which, if abolished, would cause the immediate and complete collapse of any human civilization.

Understanding practical language as technology starts with the everyday and elite working worldviews. Practical people everywhere share a common consensus everyday working worldview. Myriad elite working worldviews are specialized from this common working worldview.

These working worldviews are collections of working mindsets, models, and methods. These provide shared consensus models of how the world works and how to work in the world.

This everyday working worldview is highly pluralistic. It supports many alternative and complementary practical reality perspectives.

This pluralism of overlapping practical perspectives is highly redundant but hardly superfluous. It is exactly this pluralism that provides the power and precision of the common working worldview.

Where does this working worldview come from? We all contribute to the common everyday working worldview consensus. It is the massive collective competence of practical people everywhere that serves to order and optimize the common consensus everyday working worldview.

The resulting mass consensus everyday working worldview is perfect. It is perfect in the same sense that mass markets are perfect. It is perfect not in the universal philosophical sense but rather in the utilitarian practical sense. It is perfect in the sense of being engineering and economic optimal. So it is perfect for all practical purposes.

The perfection of the mass consensus working worldview is maintained by a dynamic dialectic. This dialectic has been underway since antiquity. Today this dialectic continues around the world around the clock on the Internet.

The starting point for deconstructing working worldviews is consensus context structure. Every sort of practical modeling is done in the context of one or more modeling communities.

The context of a modeling community is one or more community consensus modeling spaces. It is the lexical and logical conventions of this context that support interchange of practical model meaning among community authors and audiences.

A context framework serves as a sort of a map of a modeling space. A commonplace restaurant menu is an example of simple context framework for constructing a restaurant meal model from a combinatorial constructive context space of alternative meal structures.

Practical information contexts are just one of many sorts of mass consensus systems found in modern industrial society. Mass markets are the sort of mass consensus systems that are best understood today. Diverse sorts of market research techniques are the means by which we routinely deconstruct financial, consumer, commercial and other sorts of mass markets.

Minor specializations of standard market research methodologies of this sort are the state of the art in understanding practical language as technology today. These provide the tools and techniques required to deconstruct practical modeling systems, standards, spaces, and stocks.

Deconstructions of this sort are the starting point for principled practical model automation. The context elements of everyday and elite working worldviews can be found and framed by this means.

Principled practical model automation is prerequisite to any manageable means of automated principled software production. It is by this means that the feedstock for synthetic software can be conveniently prepared. It is by this means that the path from practical prose to practical programs can be advantageously automated.

## Practical Logics are State Logics, Not Set Logics

Software customers see computers as machine tools of mass modeling and mass management. Computers are seen as marvelous mathematical machines in the software industry.

These two views embody very different ideas about the nature of practical logic. Software customers see the logic of practical information as state logics. This is consistent with the view that practical information is civilization in disguise.

In the software industry it has always been assumed that the native logics of practical information are state logics. This is consistent with the view that information is computation in disguise.

State logics and set logics are very different kinds of logics. State logics are the native logics of materialist meaning. Set logics are the native logics of mathematical meaning. These two antithetical sorts of logics have very little in common.

Which sort of logics is really the right sort? Today set logics are the starting point for pattern programming. Set logics are the native logics of software languages including both code and query languages.

Yet state logics are the native logics of practical information and practical language. These are easily discovered in all sorts of practical models. Set logics are nowhere to be found.

For fifty years software science researchers have struggled to perfect practical set logics. These efforts so far have been fruitless and futile.

The reason for this is that set logics intrinsically just aren't very practical. They are just simply the wrong sort of logics for most practical purposes.

There is no shortage of practical set logic candidates. These include traditional and non-traditional logics. Some examples include modal, linear, sequent, universal, higher order, dynamic, sentential, and non-monotonic logics.

It is not impossible to model practical language or practical information in terms of set logics. Even so, this is just variant pattern design science modeling. The usual pattern problems, such as safety and scalability, prevent demonstrations from maturing into deployments.

The practical logics encoded by practical language are differential state dynamics logics. These are the sorts of practical logics that practical people understand and use everyday. Elite forms of differential state dynamics logics are employed in general management, the principled professions, and the statefull sciences.

These materialist state logics are very different from the mathematical set logics employed in commonplace programming languages. In practical computing we always use explicit set logics for encoding. Even so, it is implicit state logics that always carry all the expository practical meaning of practical programs.

We are all masters of many hundreds of specialized state logics. Some of us manage to master many thousands. There are the state logics for telephones and traffic laws. There are state logics for bookkeeping and bowling. There are state logics for cooking and chemistry. There are state logics for airframes and atomic physics.

Differential state dynamics logics are the native logics of practical reality. The intrinsic order of physical reality is solely state dynamical. The emergent order of practical reality is reducible to complex physical differential dynamics. We conveniently live in a universe where the differential meaning of intrinsic order is necessary and sufficient for all practical purposes.

## Software Science is Still a Functional Mythology Today

A mature design discipline is one which has achieved a fully mature functional mechanics. A mature functional mechanics involves a mature statefull science discipline as well as a mature empirical engineering discipline. These must be complementary. Theory and technology must go hand in hand.

Today software still has a very long way to go in order to reach this ultimate destination. Current practical software production is based on a mature functional mythology.

Functional mythology is what you do in order to make things work while you're waiting for a functional mechanics. The functional mythology of practical software production works pretty well today. Even so, it is not a substitute for a full functional mechanics of practical software production.

There's nothing wrong with functional mythology. This is how technologies grow up. Many of the major technologies we all use everyday started out with decades of functional mythology building. The steam engine, the telephone, the telegraph, electricity, and radio all started out with elaborate functional mythologies.

All these mature technologies evolved into mature functional mechanics from mature functional mythologies. Maturation of a functional mythology sets the stage for the transition to a functional mechanics. This transition is the first step on the path towards the maturation of the functional mechanics.

The lack of functional mechanics in practical software production isn't a big deal. This is an open secret in the software world. Everybody sort of knows about it. Nobody ever talks much about it. Nobody ever seems to worry much about it.

Today the functional mythology of practical software production consists of two main parts. Theses parts include a vast body of tradition plus a vast body of theory. These parts aren't very complementary.

Why is there no true technology of practical software production today? This is hardly due to lack of effort.

For fifty years software practitioners have labored to turn software tradition into true technology. For fifty years software professors have labored to turn software theory into true technology.

All of these efforts always have and always will be in vain. Prevailing pattern design science is simply inadequate to support any true technology of software production. Principled design science is necessary and sufficient for this purpose.

This lack of a true technology of practical software production explains why pattern programming is inherently recondite and risky. Few routine software design judgments can be justified today by appealing to theoretical foundations. Justifications of this sort are routine and often required in all other major design disciplines.

Today design decisions in practical software production are based mostly on pattern precedent traditions. We do what we do because it worked last time. With a bit of tinkering it might work just a bit better this time. This is how software design decisions have always been made.

Why are software design decisions made this way? Because practical software has not yet made the transition to functional mechanics. So in a very real sense we don't know what we're doing in the practical software world today.

Even so, we don't really care. We do it anyway. We do a lot of it. We have lots of fun and live well by doing it. We do it by producing lots of really cool pattern programs.

This all works remarkably well. At least it works remarkably well for present software industry purposes just as it has in the past.

Yet this works far less well for software industry customers. These customers would much prefer automated production of complex custom software. They would like to advantageously automate most of the mass modeling and mass management systems that they depend on.

This will require a full functional mechanics of practical software production. This cannot be a functional mechanics based on pattern design science. It must be a functional mechanics based on principled design science.

## Conclusions

Over the course of the last fifty years the software industry has put all of its eggs in the pattern software design science basket. The industry has frantically, furiously, and perhaps even fanatically focused on pushing pattern program production towards perfection. The industry has never seriously considered alternative principled design science approaches to practical software production.

Patten programming allows us to routinely do things with practical software that were possible only in science fiction a few decades ago. This is the brilliant success of pattern programming.

Yet practical design science tells us definitively that pattern programming has its limitations. It cannot support "smart" software. It cannot support automated mass custom complex practical software production.

These are solutions that the software industry can, must, and will eventually provide for its customers. Yet the great success of pattern programming has become a barrier to moving these future solutions forward. There is little immediate incentive to hurry these solutions along.

These solutions will require alternative principled software production approaches. Such approaches appear to be possible and practical today. They will likely be highly profitable as well.

Yet these approaches will be a radical departure from pattern programming. They will very different from software production as it has been done up until now.

These approaches will be exotic and expensive. Even so they will provide vast new opportunities for delivering software customer value.

In this and many other ways software remains an unfinished revolution. Sometimes today software can comfortably seem like a successful and settled revolution.

This is exactly why incredibly dangerous software ideas are indispensable. These serve to remind us that we have only just begun to scratch the surface of what can, must, and eventually will be done with practical software.